\documentclass[aps,pra,showpacs,superscriptaddress,twocolumn,final,numerical,10pt]{revtex4-1}
\usepackage{epsfig}
\usepackage{graphicx}
\usepackage{dcolumn}
\usepackage{bm}
\usepackage{amsthm}
\usepackage{amsfonts}
\usepackage{amscd}
\usepackage{amsmath}  
\usepackage{amssymb}
\usepackage{enumerate}
\usepackage{color}

\newcommand{\ket}[1]{\mbox{$ | #1 \rangle $}}
\newcommand{\bra}[1]{\mbox{$ \langle #1 | $}}

\begin{document}

\title{Loss-tolerant quantum cryptography with imperfect sources}

\author{Kiyoshi Tamaki}
\email{tamaki.kiyoshi@lab.ntt.co.jp}
\affiliation{NTT Basic Research Laboratories, NTT Corporation, 3-1,Morinosato Wakamiya Atsugi-Shi, Kanagawa, 243-0198, Japan}
\author{Marcos Curty}
\affiliation{EI Telecomunicaci\'on, Dept. of Signal Theory and Communications, University of Vigo, E-36310, Spain}
\author{Go Kato}
\affiliation{NTT Communication Science Laboratories, NTT Corporation, 3-1,Morinosato Wakamiya Atsugi-Shi, Kanagawa, 243-0198, Japan}
\author{Hoi-Kwong Lo}
\affiliation{Center for Quantum Information and Quantum Control, Dept. of Electrical \& Computer Engineering and Dept. of Physics,
University of Toronto, M5S 3G4, Canada}
\author{Koji Azuma}
\affiliation{NTT Basic Research Laboratories, NTT Corporation, 3-1,Morinosato Wakamiya Atsugi-Shi, Kanagawa, 243-0198, Japan}

\date{\today}

\begin{abstract}
In principle, quantum key distribution (QKD)
offers unconditional security based on the laws of physics. In
practice, flaws in the state preparation undermine the security of QKD systems, as 
standard
theoretical approaches to deal with state preparation flaws are not
loss-tolerant. An eavesdropper can enhance
and exploit such imperfections through quantum channel loss, thus
dramatically lowering the key generation rate. Crucially, the security
analyses of most existing QKD experiments are rather
unrealistic as they typically neglect this effect. 
Here, we propose a novel and general approach that makes QKD loss-tolerant to state preparation flaws. Importantly, 
it suggests that the state 
preparation process in QKD can be significantly less precise than initially thought. Our method
can widely apply to other quantum cryptographic protocols.
\end{abstract}

\pacs{03.67.Dd, 03.67.-a}
\maketitle

\textit{Introduction.}---
Quantum key distribution (QKD)~\cite{review} allows two distant parties, Alice and Bob, to distribute a secret key, which is essential 
to achieve provable secure communications~\cite{vernam}. 
The field of QKD has progressed very rapidly over the 
last years, and it now offers practical systems that can operate in realistic environments~\cite{Tokyo,NEC}.

Crucially, QKD provides unconditional security based on the laws of physics, {\it i.e.}, 
despite the computational power of the eavesdropper, Eve. Indeed, the security of QKD has been promptly demonstrated for different scenarios~\cite{sec1,sec2,sec3,sec4,sec5,sec6,Phase error,comp}. Importantly, Gottesman, Lo, L\"utkenhaus and Preskill~\cite{GLLP} 
(henceforth referred to as GLLP) 
proved the security of QKD when Alice's and Bob's devices are flawed, as is the case in practical implementations.
Unfortunately, however,
GLLP has a severe limitation,
namely, it is not loss-tolerant; it assumes 
the worst case scenario where Eve can 
enhance flaws in the state preparation by exploiting
channel loss. As a result,
the key generation rate and achievable distance of QKD 
are dramatically reduced~\cite{phaseMDI}.
Notice that most existing QKD experiments simply ignore state preparation imperfections
in their key rate formula, which
renders their results unrealistic and not really secure.

In this Letter, we show that GLLP's worst
case assumption is far too conservative, {\it i.e.}, in sharp contrast to
GLLP, we present a security proof for QKD that is loss-tolerant.
Indeed, for the case of
modulation errors, an important flaw in real-life QKD systems, we show that Eve {\it cannot}
exploit channel loss to 
enhance such imperfections. The intuition here is rather simple: in this type of state
preparation flaws the signals sent out by Alice are still qubits, {\it i.e.},
there is no side-channel for Eve to exploit to enhance
the imperfections through channel loss.

Our work
builds on the security proof introduced by Koashi~\cite{comp} based  
on complementarity of conjugate observables, X and Z. Also, it employs the idea of ``rejected data analysis''~\cite{barnett}, {\it i.e.}, we consider data obtained when Alice's and Bob's measurement bases are different, as well as the fact that any qubit state can be written in terms of Pauli matrices. Therefore, to calculate the objective quantity, {\it i.e.}, the
so-called phase error rate, it is enough to find the transmission rates of these matrices.

In so doing, we can: (i)
dramatically improve the key rate and achievable distance of QKD with modulation errors (see Fig.~1 for details);  (ii) show
that the three-state scheme~\cite{three_state,Fred} gives precisely the same key rate as the BB84 protocol~\cite{bb84}. This result is outstanding, as it implies that one of the 
signals sent in BB84 is actually redundant~\cite{wang_note}. 
In addition, our technique is: (iii) applicable to measurement-device-independent QKD (mdiQKD)~\cite{MDI}; (iv) applicable to other QKD schemes including the six-state protocol~\cite{sixstate}. It can be shown, for instance, that a particular four-state scheme can 
post-process its data following the specifications of the six-state protocol~\cite{sixstate}. That is, it 
can use the correlation between phase and bit errors to increase its key rate. 
(v) Our method also
applies to other quantum cryptographic applications ({\it e.g.}, bit commitment based on the noisy storage model~\cite{bit_commitment}).

To simplify the discussion, we assume collective attacks, {\it i.e.}, Eve applies the same quantum operation to each signal. 
However, our results also hold against coherent attacks by just applying either the quantum De Finetti theorem~\cite{defi} or Azuma's inequality~\cite{Azuma0, Azuma1, Azuma2} (see Appendix A for details). 
Moreover, for simplicity, we consider the asymptotic scenario where Alice sends Bob an infinite number of signals. 
In addition, we assume that there is no side-channel in the source. 
That is, we consider that the single-photon components of Alice's signals are qubits,
and we analyze an important type of state preparation flaws,
namely modulation errors due to
slightly over or under modulation of the signal's phase/polarization
by an imperfect apparatus.
Also, we assume that 
Bob's measurement device satisfies two conditions: {\it random basis choice} and {\it basis-independent detection efficiency}. 
The former is fulfilled if Bob 
selects at random between two or more measurement settings; one for key distillation and the others for parameter estimation. 
The latter is satisfied if the probability of having a detection event is independent of Bob's measurement setting choice. 
With mdiQKD, we can waive these
two conditions and allow the detection system to be untrusted.

\textit{Prepare\&measure three-state protocol.}---In this 
scheme~\cite{three_state,Fred}, Alice sends Bob three pure states,
$\ket{\phi_{0{\rm z}}}=\ket{0_{\rm z}}$, $\ket{\phi_{1{\rm z}}}=\ket{1_{\rm z}}$ and $\ket{\phi_{0{\rm x}}}=\ket{0_{\rm x}}$, 
which she selects independently at random for each signal. Here, the 
states $\ket{j_{\rm x}}=[\ket{0_{\rm z}}+(-1)^j\ket{1_{\rm z}}]/\sqrt{2}$, 
with $j\in\{0,1\}$.
On Bob's side, 
he measures the signals received using either the X or the Z basis, which he selects as well independently at random
for each incoming signal. After that, 
Alice and Bob announce their basis choices, and they estimate the bit and phase error rate. 
We assume that they generate a secret key only from those instances where 
both of them select say the Z basis. 

In the following, we present a precise phase error rate estimation technique that uses 
the bases mismatch events information. The key idea is very simple yet 
potentially very useful: since any qubit state can be written in terms of Pauli matrices, it is enough to find the 
transmission rates of these operators; this will become clear below. First, we
introduce some notation. 

In particular, let $\{{\hat M_{0\beta}}, {\hat M_{1\beta}}, {\hat M_{\rm f}}\}$ denote the elements of Bob's 
positive-operator valued measure (POVM) associated with the basis $\beta\in\{{\rm X,Z}\}$. 
${\hat M_{0\beta}}$ and ${\hat M_{1\beta}}$ correspond, respectively, to the bit values $0$ and $1$, and
${\hat M_{\rm f}}$ represents the inconclusive event.
These operators 
do not necessarily act on a qubit space, {\it i.e.}, Eve can send Bob any higher-dimensional state. 
The essential assumption here is that ${\hat M}_{\rm f}$ is the same for both bases \cite{comp}. 
Also, we denote as $Y_{s_{\beta}, j_{\alpha}}$, with $s,j\in\{0,1\}$ and $\beta,\alpha\in\{{\rm X,Z}\}$, the joint 
probability that Alice prepares the state $\ket{\phi_{j\alpha}}$ and Bob measures 
it in the $\beta$ basis and obtains a bit value $s$. 

\noindent{\it Theorem. The prepare\&measure three-state protocol described above 
provides a secret key rate $R\propto1-h(e_{\rm z})-h(e_{\rm x})$, where 
$h(x)=-x\log_{2}x-(1-x)\log_{2}(1-x)$ is the binary Shannon entropy, $e_{\rm z}$ is the bit error rate, 
and $e_{\rm x}$ is the phase error rate given by
\begin{equation}\label{eq_thm}
e_{\rm x}=\frac{Y_{0_{\rm x},0_{\rm z}}+Y_{0_{\rm x},1_{\rm z}}+Y_{1_{\rm x},0_{\rm x}}-Y_{0_{\rm x},0_{\rm x}}}
{Y_{0_{\rm x},0_{\rm z}}+Y_{0_{\rm x},1_{\rm z}}+Y_{1_{\rm x},0_{\rm z}}+Y_{1_{\rm x},1_{\rm z}}}.
\end{equation}
Notably, $e_{\rm x}$ coincides with that of the BB84 protocol.}

This result is remarkable because it implies that the three-state protocol can achieve precisely the 
same performance as the BB84 scheme, since both protocols can obtain the {\it exact} value for the phase error rate $e_{\rm x}$ 
together with the bit error rate $e_{\rm z}$. That is, 
the additional signal $\ket{1_{\rm x}}$ that is sent in BB84 seems to be unnecessary. This means, for instance, that in
those implementations of the BB84 protocol that use four laser sources one could keep one laser just as back-up in case one 
of them fails, without any decrease in performance \cite{Free-space}.
This also reduces the consumption of random numbers to select the different sources.
 Our security analysis differs from that
provided in Ref.~\cite{Fred} in that it
requires less privacy amplification (PA), and, consequently, it can deliver a higher secret key rate. Next, we present the proof 
for the Theorem. 

\noindent{\it Proof.} 
The preparation of the Z-basis states $\ket{\phi_{0{\rm z}}}$ and $\ket{\phi_{1{\rm z}}}$ 
can be formulated in an entanglement based version of the protocol as follows.
Alice first creates a source state 
$\ket{\Psi_{\rm Z}}_{\rm AB}=(\ket{0_{\rm z}}_{\rm A}\ket{\phi_{0{\rm z}}}_{\rm B}+\ket{1_{\rm z}}_{\rm A}\ket{\phi_{1{\rm z}}}_{\rm B})/\sqrt{2}$. Afterwards,
she measures system $\rm A$ in the Z basis, thereby producing the correct signal state at site $\rm B$ that is sent to Bob.  
The phase error rate $e_{\rm x}$ is defined as the bit error rate that Alice and Bob would observe if they 
measure $\ket{\Psi_{\rm Z}}_{\rm AB}$ in the X basis. 
Importantly, if 
$N_{\rm z}$ denotes the number of sifted bits in the Z basis,  
to distill a secure key Alice and Bob need to sacrifice $N_{\rm z}h(e_{\rm x})$ bits in the PA step. 

To calculate 
$e_{\rm x}$, we define a virtual protocol where Alice and Bob measure
$\ket{\Psi_{\rm Z}}_{\rm AB}$ in the X basis.
This state can be equivalently written as
$\ket{\Psi_{\rm Z}}_{\rm AB}=(\ket{0_{\rm x}}_{\rm A}\ket{0_{\rm x}}_{\rm B}+\ket{1_{\rm x}}_{\rm A}\ket{1_{\rm x}}_{\rm B})/\sqrt{2}$. 
That is, if Alice measures system $\rm A$ in the X basis and obtains the 
bit value $j\in\{0,1\}$, she effectively prepares the signal
$\ket{j_{\rm x}}_{\rm B}$ at site $\rm B$. 
This means that
$e_{\rm x}=(Y_{0_{\rm x},1_{\rm x}}+Y_{1_{\rm x},0_{\rm x}})/(Y_{0_{\rm x},0_{\rm x}}+Y_{1_{\rm x},0_{\rm x}}+Y_{0_{\rm x},1_{\rm x}}+Y_{1_{\rm x},1_{\rm x}})$. Importantly, the probabilities $Y_{s_{\rm x},0_{\rm x}}$, with
$s\in\{0,1\}$, are directly observed in the experiment because in the 
actual protocol Alice sends Bob the signal $\ket{0_{\rm x}}$. 

To obtain the terms $Y_{s_{\rm x},1_{\rm x}}$ we use the fact that any qubit state
can be decomposed in terms of the identity and the three Pauli matrices. For this, we first rewrite 
$Y_{s_{\rm x}, 1_{\rm x}}=\frac{1}{6}{\rm Tr}[{\hat D}_{s_{\rm x}}{\hat P}(\ket{1_{\rm x}})]$, where ${\hat P}(\ket{\phi})=\ket{\phi}\bra{\phi}$, 
${\hat D}_{s_{\rm x}}= \sum_{k}{\hat A}_{k}^{\dagger}{\hat M_{s_{\rm x}}}{\hat A}_{k}$ with ${\hat A}_{k}$ being an arbitrary 
operator (see Appendix A), and $1/6$ is the probability that Alice emits $\ket{1_{\rm x}}$ and Bob chooses the X basis. 
Then, we define $q_{s_{\rm x}| t}={\rm Tr}({\hat D}_{s_{\rm x}}{\hat \sigma_{t}})/2$ where  
${\hat \sigma_{t}}$, with $t\in\{{\rm Id},x,z\}$, denotes, respectively, the identity and two of the Pauli operators. With 
this notation, and using ${\hat P}(\ket{1_{\rm x}})=({\hat \openone}-{\hat \sigma_{x}})/2$,
we have that $Y_{s_{\rm x}, 1_{\rm x}}=\frac{1}{6}(q_{s_{\rm x}| {\rm Id}}-q_{s_{\rm x}| x})$. 
Finally, to calculate  
$q_{s_{\rm x}| {\rm Id}}$ and $q_{s_{\rm x}| x}$ we use the following constraints,
\begin{eqnarray}
Y_{s_{\rm x}, 0_{\rm z}}&=&\frac{1}{6}{\rm Tr}\left[{\hat D}_{s_{\rm x}}{\hat P}(\ket{\phi_{0{\rm z}}})\right]=\frac{1}{6}(q_{s_{\rm x}| {\rm Id}}+q_{s_{\rm x}| z}),\label{L1} \\
Y_{s_{\rm x}, 1_{\rm z}}&=&\frac{1}{6}{\rm Tr}\left[{\hat D}_{s_{\rm x}}{\hat P}(\ket{\phi_{1{\rm z}}})\right]=\frac{1}{6}(q_{s_{\rm x}| {\rm Id}}-q_{s_{\rm x}| z}),\label{L2} \\
Y_{s_{\rm x}, 0_{\rm x}}&=&\frac{1}{6}{\rm Tr}\left[{\hat D}_{s_{\rm x}}{\hat P}(\ket{\phi_{0{\rm x}}})\right]=\frac{1}{6}(q_{s_{\rm x}| {\rm Id}}+q_{s_{\rm x}| x}).\label{L3} 
\end{eqnarray}
Recall that the probabilities $Y_{s_{\rm x}, 0_{\rm z}}$, $Y_{s_{\rm x}, 1_{\rm z}}$ and $Y_{s_{\rm x}, 0_{\rm x}}$ are directly measured 
in the experiment. Also, we have that Eqs.~(\ref{L1})-(\ref{L3}) are independent, since
the vectors ${\vec V_{j\alpha}}:=(1, p_{x}^{j\alpha}, p_{y}^{j\alpha}, p_{z}^{j\alpha})$, 
with $p_{w}^{j\alpha}$ being the $w$ $(={\rm x}, {\rm y}, {\rm z})$ component of the Bloch vector of the state $\ket{\phi_{j\alpha}}$, are mutually linearly independent. 
Thus, by solving Eqs.~(\ref{L1})-(\ref{L3})
one can obtain the {\it exact} value for 
$q_{s_{\rm x}|t}$; we find that $Y_{s_{\rm x}, 1_{\rm x}}=Y_{s_{\rm x}, 0_{\rm z}}+Y_{s_{\rm x}, 1_{\rm z}}-Y_{s_{\rm x}, 0_{\rm x}}$.
Substituting this expression into the definition of $e_{\rm x}$ we obtain Eq.~(\ref{eq_thm}).
$\square$

So far, for simplicity, we have considered that Alice sends Bob single-photon states. 
However, our results can be used as well when she
prepares phase-randomized weak coherent pulses (WCPs) in combination with decoy states~\cite{Decoy}. This is so 
because the decoy-state method allows Alice and Bob to estimate the relevant probabilities  
$Y_{s_{\rm x}, 0_{\rm z}}$, $Y_{s_{\rm x}, 1_{\rm z}}$, and $Y_{s_{\rm x}, 0_{\rm x}}$ associated with the single-photon signals. 
In addition, the analysis above can be easily extended to include modulation errors (see Appendix B). 
This scenario is shown in the simulation. 

\begin{figure}
\begin{center}
 \includegraphics[scale=0.4]{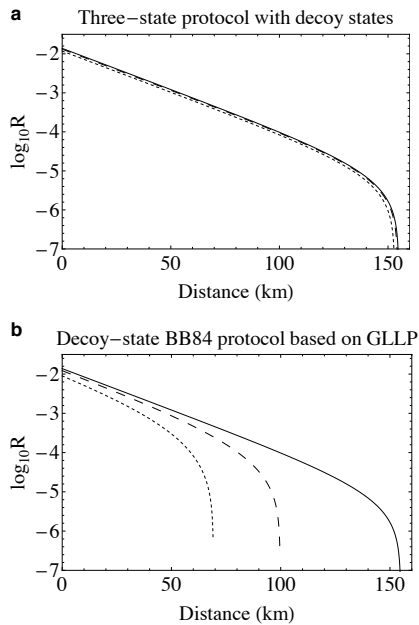}
 \end{center}
 \caption{Lower bound on the secret key rate $R$ for different values 
of the phase modulation error $\delta$. (a) Three-state protocol with WCPs and decoy states. Here we use the 
 phase error rate estimation technique introduced in the paper. (b) Decoy-state BB84 protocol based on the GLLP security argument~\cite{GLLP}.  
 For simulation purposes, we consider the following experimental parameters:
 the dark count rate of Bob's detectors is $0.5\times10^{-7}$, 
 the overall transmittance of his detection apparatus is $0.15$, the loss coefficient of the channel is
 $0.21$ dB/km, and the efficiency of the error correction protocol is $1.22$. 
The solid, dashed, and dotted lines correspond, respectively, to the cases $\delta=0, 0.063$ and $0.126$. The case
$\delta\ge0.063$ corresponds to an experimentally available value~\cite{PM}. For each line, we optimise the intensity of the
signals to maximise the key rate. The solid lines in both figures coincide. 
Importantly, in (a) the three lines almost overlap, {\it i.e.}, Eve cannot enhance state preparation flaws by exploiting the channel loss. 
This shows a dramatic improvement over the results illustrated in (b) based on the previous technique (in GLLP).
\label{comparison}}
\end{figure} 
\textit{Simulation.}---
Here we evaluate 
the performance of a three-state protocol based on WCPs together with decoy states in the presence of modulation errors.
For simplicity, we consider the asymptotic situation where Alice uses an
infinite number of decoy settings. Moreover, 
we assume that 
she employs phase-coding, as this is usually the preferential coding choice in optical fibre implementations. 
However, our analysis applies as well to other coding schemes, {\it e.g.}, polarization and time-bin coding. 

More precisely, we consider that Alice sends Bob
signals of the form $\ket{e^{i\xi}\sqrt{\alpha}}_{r}\ket{e^{i(\xi+\theta_{\rm A}+\delta\theta_{\rm A}/\pi)}\sqrt{\alpha}}_{s}$, where $\xi\in[0,2\pi)$ 
is a random phase, $\theta_{\rm A}\in\{0,\pi/2,\pi\}$ encodes Alice's information, 
the term $\delta\theta_{\rm A}/\pi$ with $\delta\ge0$ models an example of phase modulation errors,
and 
$\ket{e^{i\xi}\sqrt{\alpha}}_{r}$ is a coherent state with mean photon number $\alpha$. 
The subscripts $r$ and $s$ are used to denote, respectively, the reference and signal mode. In this scenario, 
the single-photon components of Alice's signals lie on a plane of the Bloch sphere. 
In addition, we assume the same phase modulation error 
on Bob's side, {\it i.e.}, his phase modulation is $\theta_{\rm B}+\delta\theta_{\rm B}/\pi$ when he 
chooses $\theta_{\rm B}\in\{0,\pi/2\}$. Importantly, since $\delta\ge0$, Alice's and Bob's modulation errors do
not cancel each other, but they only increase the total modulation error. 

The resulting lower bound on the secret key rate $R$ for different values 
of the error parameter $\delta$ is shown in Fig.~\ref{comparison} (see Appendix C). For comparison, this figure includes as well a lower bound on $R$ for
the asymptotic decoy-state BB84 protocol. For the latter, we use results from Ref.~\cite{phaseMDI}, which are based on the GLLP security 
analysis~\cite{GLLP}, and we use the same phase modulation model described above with $\theta_{\rm A}\in\{0,\pi/2,\pi,3\pi/2\}$. As shown
in the figure, our phase error rate estimation technique can significantly outperform GLLP in the presence of modulation errors.  
In particular, while GLLP delivers a key rate that decreases rapidly when $\delta$ increases (since it considers the worst case scenario where 
losses can increase 
the fidelity flaw~\cite{phaseMDI}), our method produces an almost constant key rate
 independently of $\delta$. The slight performance decrease of the three-state protocol when 
$\delta$ increases is due to 
the increase of the bit error rate $e_{\rm z}$ stemming from imperfect phase modulations.

\textit{Measurement-device-independent QKD.}---
We consider a modified version of mdiQKD~\cite{MDI} where Alice and Bob send Charles  
the states 
$\ket{\phi_{0{\rm z}}}$, $\ket{\phi_{1{\rm z}}}$, and $\ket{\phi_{1{\rm x}}}$.
Charles is supposed to perform a Bell state measurement that projects them into a Bell state, and then he
announces his results. 
Alice and Bob keep the data associated with the successful results, post-select the events where they employ the same basis, and  
say Bob applies a bit flip to part of his data~\cite{MDI}.
They use the Z basis (X basis) for key distillation (parameter estimation). 

In the following, we apply the phase error rate estimation method introduced above to mdiQKD. Now, $e_{\rm x}$ can be
expressed as
\begin{equation}
e_{\rm x}=\frac{Y_{\phi^+, 0_{\rm x}1_{\rm x}}+Y_{\phi^+, 1_{\rm x}0_{\rm x}}}
{Y_{\phi^+, 0_{\rm x}1_{\rm x}}+Y_{\phi^+, 1_{\rm x}0_{\rm x}}+Y_{\phi^+, 0_{\rm x}0_{\rm x}}+Y_{\phi^+, 1_{\rm x}1_{\rm x}}}, 
\end{equation}
where $Y_{\phi^+, j_{\rm x}k_{\rm x}}$, with $j,k\in\{0,1\}$, denotes the joint probability that Alice and Bob send 
Charles $\ket{j_{\rm x}}$ and $\ket{k_{\rm x}}$ respectively, and Charles declares the result
$\ket{\phi^+}$ (although he might be dishonest). This probability can be expressed as
$Y_{\phi^+, j_{\rm x}k_{\rm x}}=\frac{1}{9}{\rm Tr}[{\hat D}_{\phi^+}{\hat P}(\ket{j_{\rm x}})\otimes{\hat P}(\ket{k_{\rm x}})]$ for a certain
operator ${\hat D}_{\phi^+}$. Now, we follow the technique described previously. 
We define 
$q_{\phi^+| s,t}={\rm Tr}({\hat D}_{\phi^+}{\hat \sigma_{s}}\otimes{\hat \sigma_{t}})/4$ with
$s,t\in\{{\rm Id},x,z\}$, and we use 
${\hat P}(\ket{j_{\rm x}})=[{\hat \openone}+(-1)^{j}{\hat \sigma_{x}}]/2$ to write
$Y_{\phi^+, j_{\rm x}k_{\rm x}}$ in terms of
$q_{\phi^+| {\rm Id},{\rm Id}}$, $q_{\phi^+| x,{\rm Id}}$, $q_{\phi^+| {\rm Id},x}$, and $q_{\phi^+| x,x}$. Finally, to calculate these coefficients we solve the following set of linear equations,
\begin{eqnarray}
Y_{\phi^+, j_{\rm z}k_{\rm z}}&=&\frac{\gamma}{9}
{\rm Tr}\left[{\hat D}_{\phi^+}{\hat P}(\ket{\phi_{j{\rm z}}})\otimes{\hat P}(\ket{\phi_{k{\rm z}}})\right],\ \label{testneeded}\\
Y_{\phi^+, 0_{\rm x}k_{\rm z}}&=&\frac{1}{9}{\rm Tr}\left[{\hat D}_{\phi^+}{\hat P}(\ket{\phi_{0{\rm x}}})\otimes{\hat P}(\ket{\phi_{k{\rm z}}})\right],\label{testneeded2}\\
Y_{\phi^+, j_{\rm z}0_{\rm x}}&=&\frac{1}{9}{\rm Tr}\left[{\hat D}_{\phi^+}{\hat P}(\ket{\phi_{j{\rm z}}})\otimes{\hat P}(\ket{\phi_{0{\rm x}}})\right],\label{testneeded3}\\
Y_{\phi^+, 0_{\rm x}0_{\rm x}}&=&\frac{1}{9}{\rm Tr}\left[{\hat D}_{\phi^+}{\hat P}(\ket{\phi_{0{\rm x}}})\otimes{\hat P}(\ket{\phi_{0{\rm x}}})\right].\label{testneeded4}
\end{eqnarray}
For simplicity, here we have omitted the explicit dependence of Eqs.~(\ref{testneeded})-(\ref{testneeded4})
with $q_{\phi^+| s,t}$, and 
$\frac{\gamma}{9}$ ($0<\gamma<1$)
is the probability that 
Alice and Bob send 
Charles $\ket{\phi_{j{\rm z}}}$ and $\ket{\phi_{k{\rm z}}}$
respectively, and they sacrifice such
instances as test bits. 
Unlike the three-state protocol introduced above, 
note that now
we need such test bits from the sifted bits in the Z basis to estimate $e_{\rm x}$.
Importantly, since 
the set of vectors ${\vec V_{j\alpha}}$ associated with the states ${\hat P}(\ket{\phi_{j\alpha}})$ are mutually linearly independent,
Eqs.~(\ref{testneeded})-(\ref{testneeded4}) are also independent. Therefore,
one can obtain the exact value for all $q_{\phi^+| s,t}$ and, consequently, also for $Y_{\phi^+, j_{\rm x}k_{\rm x}}$ and $e_{\rm x}$. 

Like the three-state protocol, the mdiQKD scheme above is also loss-tolerant to modulation errors. That is, by combining our
work with mdiQKD, we can simultaneously address flaws in state
preparation and detection systems and obtain a high secret key rate.

\textit{Discussion.}---
To find the phase error rate in a QKD protocol one has to estimate the transmission rate of certain states, which might not have been sent in the 
actual scheme ({\it e.g.}, the signal $\ket{1_{\rm x}}$ in the three-state protocol), based on the observed data. 
As any qubit state can be written in terms of 
the identity and Pauli matrices, it is enough to find the 
transmission rates of these operators. In the three-state scheme, the states $\ket{\phi_{0{\rm z}}}$ and $\ket{\phi_{1{\rm z}}}$ give the 
transmission rate of ${\hat \openone}$ and ${\hat \sigma_{z}}$. Thus, by sending {\it any} other state 
on the X-Z plane of the Bloch sphere ({\it e.g.}, the signal $\ket{0_{\rm x}}$) one can determine the transmission rate of ${\hat \sigma_{x}}$ and, 
consequently, of any qubit state in that plane, including $\ket{1_{\rm x}}$. In general, 
we have that as long as the terminal points of the Bloch vectors of the three
states form a triangle it is always possible to estimate $e_{\rm x}$ precisely (see Appendix A).

Similarly, if Alice sends Bob four different qubit states, 
whose vectors ${\vec V_{j\alpha}}$ are mutually linearly independent, {\it i.e.}, the terminal points of their Bloch vectors form a triangular pyramid, one can obtain the 
transmission rate of any Pauli operator, including the identity matrix, and, therefore, also the exact transmission rate of {\it any} qubit state (see Appendix A). 
In the original mdiQKD scheme~\cite{MDI}
this implies, for instance, 
that Alice and Bob could determine the bit error rate associated to the 
virtual state that they would generate when measuring the first subsystem of $\ket{\Psi_{\rm Z}}=(\ket{0_{\rm z}}\ket{\phi_{0{\rm z}}}_{\rm C}+\ket{1_{\rm z}}\ket{\phi_{1{\rm z}}}_{\rm C})/\sqrt{2}$ in the Y-basis. Here C denotes the system that is sent to Charles. As a
result, they could directly use this information to improve the achievable secret key rate. 
In standard prepare\&measure QKD protocols, however, 
the estimation of the fictitious Y-basis error rate
requires
that Bob performs a measurement in that basis (see Appendix A). In addition, the basis-independent detection efficiency assumption must hold
and Bob's POVM elements must act on a 
qubit space. This is so because in order to exploit the 
correlation Alice and Bob need to share qubit states~\cite{sixstate}. Note, however, that this last requirement could be avoided by 
using either the universal squash idea~\cite{US} or the detector-decoy method~\cite{detector decoy}. 
That is, by including an additional phase modulator on Bob's side (to perform the Y-basis measurement) one could 
enhance the performance of several practical systems that generate such four states, {\it e.g.},  
those based on the BB84 protocol~\cite{NEC} or on the coherent-one-way (COW) scheme~\cite{COW}.

We have discussed the phase error rate estimation problem, which affects the PA step of a QKD protocol. To generate a secure key, 
however, it is also important that the bit error rate, which affects the error-correction step, is small enough. In this respect, our analysis suggests that while it is important to have a
precise state preparation in the key generation basis, that of the other basis is not as essential, which
simplifies experimental implementations. For instance, with our results,
mdiQKD only needs to align one basis well and can tolerate substantial errors
in the alignment of the other bases.

Finally, we would like to emphasize that our technique requires
a complete characterization of 
the signal states transmitted \cite{Fred2}. In practice, however, it might be easier to estimate a set of states that very likely contains the signals 
prepared. In this case, one could directly apply our method by just selecting the signal from that set that 
minimizes the key rate.
Importantly, our results show that the effect of modulation errors on the performance of practical QKD systems is 
almost negligible.

\textit{Conclusion.}---
We have introduced a phase error rate estimation method that makes QKD loss-tolerant to state preparation flaws. It uses 
information from bases mismatch events. 
We have applied this technique to different practical QKD systems 
and we have shown that it can substantially improve their key generation rate and covered distance when 
compared to the standard GLLP result. Our work constitutes an important step towards secure QKD with imperfect devices.

\textit{Acknowledgments.}---
We thank C.-H. F. Fung, F. Xu, and X.-B.Wang for fruitful discussions. 
We acknowledge support from the National Institute of Information and Communications Technology (NICT) of Japan 
(project ``Secure photonic network technology'' as part of ``The project UQCC''), the Japan Society for the Promotion of Science (JSPS) through its Funding Program for World-Leading Innovative R$\&$D on Science and Technology (FIRST Program), the European Regional Development Fund (ERDF), the Galician 
Regional Government (projects CN2012/279 and CN 2012/260, ÒConsolidation of Research Units: AtlantTICÓ), NSERC, and the CRC program.

\appendix

\section{Three-state protocol \& coherent attacks}\label{A0}
Here we present the security proof for the three-state protocol. We consider that Alice and Bob distill key 
only from those events where both of them use the Z basis, while the events where Bob employs the X basis are used
for parameter estimation.

\begin{figure}
\begin{center}
 \includegraphics[scale=0.5]{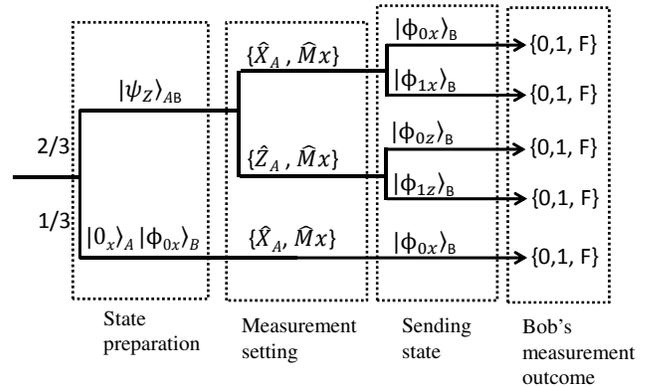}
 \end{center}
 \caption{This diagram illustrates Alice's state preparation process together with Bob's POVM ${\hat M}_{\rm x}$, the signal state sent by 
Alice (denoted as ``sending state"), and Bob's measurement outcomes. ``F'' represents the inconclusive event.
 The measurement $\{{\hat X}_{\rm A}, {\hat M}_{\rm x}\}$ on the state $\ket{\Psi_{\rm Z}}_{\rm AB}$ corresponds to the virtual protocol; 
 the phase error rate $e_{\rm x}$ is defined in terms of its outcomes.
\label{fig-appen}}
\end{figure}

As already introduced in the main text, the preparation of the Z-basis states can be 
equivalently described as follows. Alice first generates 
$\ket{\Psi_{\rm Z}}_{\rm AB}=(\ket{0_{\rm z}}_{\rm A}\ket{\phi_{0{\rm z}}}_{\rm B}+\ket{1_{\rm z}}_{\rm A}\ket{\phi_{1{\rm z}}}_{\rm B})/\sqrt{2}$, and, afterwards,
she measures system $\rm A$ in the Z basis. We denote this measurement as ${\hat Z}_{\rm A}$. Likewise, 
the preparation of the signal $\ket{0_{\rm x}}$ can also be formulated as a two-step process, {\it i.e.}, Alice first produces 
$\ket{\Psi_{\rm X}}_{\rm AB}=\ket{0_{\rm x}}_{\rm A}\ket{\phi_{0{\rm x}}}_{\rm B}$ and then she measures system A in the X basis. This measurement is denoted as ${\hat X}_{\rm A}$.
This situation is illustrated in Fig.~\ref{fig-appen}. This figure contains as well Bob's POVM ${\hat M}_{\rm x}=\{{\hat M_{0{\rm x}}}, {\hat M_{1{\rm x}}}, {\hat M_{\rm f}}\}$, together 
with his measurement outcomes. 
Note, however, that 
Fig.~\ref{fig-appen} only shows the relevant events 
that are needed to estimate the phase error rate $e_{\rm x}$.
An essential assumption here is that the operator ${\hat M_{\rm f}}$ is the same for both bases, X and Z.
Conceptually, this means that Bob could have heralded the receipt of a state
from Alice before he decides the measurement basis. This conceptual ability
to postpone the measurement basis choice is crucial for the security proof to go through. The measurement $\{{\hat X}_{\rm A}, {\hat M}_{\rm x}\}$ (on the state $\ket{\Psi_{\rm Z}}_{\rm AB}$) belongs to the virtual protocol defined in the main text; it is necessary to calculate 
$e_{\rm x}$.  

In what follows, we consider the probability distribution for the different paths in Fig.~\ref{fig-appen}. The state preparation, 
the measurement setting choice, and the selection of the state that is sent can 
be equivalently represented by
the preparation of 
\begin{eqnarray}\label{state}
\ket{\varphi}_{\rm sh,B}:=\sum_{c=1,2,3,4,5}\sqrt{P(c)}\ket{c}_{\rm sh}\ket{\phi^{(c)}}_{\rm B}\,,
\end{eqnarray}
followed by an orthogonal measurement 
using the basis $\{\ket{c}\}$ on the shield system sh possessed by
Alice. Here, the index
$c=1,2,\ldots,5$ 
identifies the five possible 
``sending states" shown in Fig.~\ref{fig-appen}. That is,
$\ket{\phi^{(1)}}_{\rm B}:=\ket{\phi_{0_{\rm x}}}_{\rm B}$,
$\ket{\phi^{(2)}}_{\rm B}:=\ket{\phi_{1_{\rm x}}}_{\rm B}$, $\ket{\phi^{(3)}}_{\rm B}:=\ket{\phi_{0_{\rm z}}}_{\rm B}$,
$\ket{\phi^{(4)}}_{\rm B}:=\ket{\phi_{1_{\rm z}}}_{\rm B}$, and $\ket{\phi^{(5)}}_{\rm B}:=\ket{\phi_{0_{\rm x}}}_{\rm B}$. 

Suppose that Alice prepares many systems of the form given by Eq.~(\ref{state}) and sends system B
to Bob through the quantum channel. Also, suppose that Alice and Bob measure in order the shield and B systems
using respectively the basis $\{\ket{c}\}$ and the POVM ${\hat M}_{\rm x}$,
and let us consider the $l^{th}$ run of the protocol.
According to Azuma's inequality, 
once we obtain the probabilities for the different paths of Fig.~\ref{fig-appen}
in the $l^{th}$ run conditioned on all previous measurement outcomes, we 
can determine the actual occurrence number of the corresponding events 
(see Refs.~\cite{Azuma1,Azuma2} for a proof of this statement). 

Next, we calculate these conditional probabilities.
For this, let 
$\ket{\Phi}_{\rm sh,B}=\ket{\varphi_{l-1}}_{\rm sh,B}\ket{\varphi_l}_{\rm sh,B}\ket{\varphi_r}_{\rm sh,B}$ denote the state 
prepared by Alice in an execution of the protocol. Here, 
$\ket{\varphi_{l-1}}_{\rm sh,B}$, $\ket{\varphi_l}_{\rm sh,B}$, and $\ket{\varphi_r}_{\rm sh,B}$ 
represent, respectively, Alice's signals in the first $l-1$ runs, in the $l^{th}$ run, and in the rest of runs. 

This state evolves according to Eve's unitary transformation, ${\hat V_{\rm BE}}$, on Bob's system $\rm B$ and on 
her system $\rm E$ as follows, 
\begin{equation}
{\hat V_{\rm BE}}\ket{\Phi}_{\rm sh,B}\ket{0}_{\rm E}
=\sum_{k}{\hat B}_{k, {\rm B}}\ket{\Phi}_{\rm sh,B}\ket{k}_{\rm E},
\end{equation}
where ${\hat B}_{k, {\rm B}}$ is a Kraus operator acting on system B. Importantly, ${\hat V_{\rm BE}}$ and 
${\hat B}_{k, {\rm B}}$ are independent of the state preparation process. This is so 
because the classical communication between Alice and Bob is done {\it after} finishing the measurements. 
Let the joint operator 
\begin{equation}
{\hat O_{l-1,{\rm sh,B}}}=\otimes_{u=1}^{l-1}{\hat M_{{\rm sh}_{u},s_{u}}},
\end{equation}
where ${\hat M_{{\rm sh}_{u},s_{u}}}$ denotes the Kraus
operator associated to the $u^{th}$ measurement outcome of the shield system
${\rm sh}$ and Bob's $u^{th}$ measurement outcome. 
The joint state in the 
$l^{th}$ run of the protocol, ${\hat\rho}_{l|O_{l-1}}^{{\rm sh,B}}$, conditioned on the measurement outcomes $O_{l-1}$
of the first $l-1$ joint systems can then be written as
\begin{eqnarray}
{\hat\rho}_{l|O_{l-1}}^{\rm sh,B}
&=&{\hat \sigma}_{l|O_{l-1}}^{\rm sh,B}/p^{(l)}\,,\\
{\hat \sigma}_{l|O_{l-1}}^{\rm sh,B}&:=&\sum_{k}{\rm Tr}_{{\overline l}}\left[{\hat P}\left({\hat O_{l-1,{\rm sh,B}}}{\hat B}_{k, {\rm B}}\ket{\Phi}_{\rm sh,B}\right)\right]\,,\nonumber\\
p^{(l)}&:=&{\rm Tr}\left({\hat \sigma}_{l|O_{l-1}}^{\rm sh,B}\right)\,,\nonumber
\label{lth}
\end{eqnarray}
where ${\rm Tr}_{\overline l}$ represents the partial trace over all 
systems except the $l^{th}$ ${\rm sh}$ and ${\rm B}$ systems. 
Equivalently, ${\hat \sigma}_{l|O_{l-1}}^{\rm sh,B}$ can be rewritten as
\begin{eqnarray}
&&{\hat \sigma}_{l|O_{l-1}}^{\rm sh,B}=\sum_{k}\sum_{{\vec x}_{l-1},{\vec x}_{r}}{\hat P}\left({\hat A}_{k,{\rm B}|O_{l-1}}^{({\vec x}_{l-1},{\vec x}_{r})}\ket{\varphi_l}_{\rm sh,B}\right) \nonumber\\
&=&\sum_{k}\sum_{{\vec x}_{l-1},{\vec x}_{r}}{\hat P}\left(\sum_{c}\sqrt{P(c)}\ket{c}_{\rm sh}{\hat A}^{({\vec x}_{l-1},{\vec x}_{r})}_{k,{\rm B}|O_{l-1}}\ket{\phi^{(c)}}_{\rm B}\right), \quad
\label{density1}
\end{eqnarray}
where
\begin{eqnarray}
{\hat A}_{k,{\rm B}|O_{l-1}}^{({\vec x}_{l-1},{\vec x}_{r})}
&:=&\bra{{\vec x}_{r}}\bra{{\vec x}_{l-1}}{\hat O_{l-1,{\rm sh,B}}}
{\hat B}_{k,{\rm B}}\ket{\varphi_{l-1}}_{\rm sh,B}\ket{\varphi_r}_{\rm sh,B}.\nonumber 
\label{kraus}
\end{eqnarray} 
Here $\{\bra{{\vec x}_{r}}\}$ ($\{\bra{{\vec x}_{l-1}}\}$) is bases for all the remaining 
systems after the $l^{th}$ run (for the first $l-1$ joint systems).
Importantly, Eq.~(\ref{density1}) states 
that the $l^{th}$ joint system is subjected to Eve's action and
her action depends on all the previous measurement outcomes on the first $l-1$ joint systems.

Now, to determine the probability distribution for the different
paths in Fig.~\ref{fig-appen}, we measure the shield system sh using the basis $\{\ket{c}\}$ and Bob's system using the 
X basis. The probability of obtaining $c$ and the bit value
$s_{\rm x}$ conditioned on $O_{l-1}$ is given by 
\begin{eqnarray}
Y_{s_{\rm x}, c|O_{l-1}}&=&\frac{P(c)}{p^{(l)}}\sum_{k}\sum_{{\vec x}_{l-1},{\vec x}_{r}}{\rm Tr}[{\hat P}({\hat A}^{({\vec x}_{l-1},{\vec x}_{r})}_{k,{\rm B}|O_{l-1}}\ket{\phi^{(c)}}_{\rm B}){\hat M_{s_{\rm x}}}]\nonumber\\
&:=&\frac{P(c)}{p^{(l)}}{\rm Tr}[{\hat D}_{s_{\rm x}|O_{l-1}}{\hat P}(\ket{\phi^{(c)}}_{\rm B})],
\label{AzumaY}
\end{eqnarray}
where 
\begin{displaymath}
{\hat D}_{s_{\rm x}|O_{l-1}}=\sum_{k}\sum_{{\vec x}_{l-1},{\vec x}_{r}}{\hat A}^{\dagger({\vec x}_{l-1},{\vec x}_{r})}_{k,{\rm B}|O_{l-1}}{\hat M_{s_{\rm x}}}{\hat A}_{k,{\rm B}|O_{l-1}}^{({\vec x}_{l-1},{\vec x}_{r})}.
\end{displaymath}
Note that the discussions in the main text use the probability $P(c)$ given in Eq.~(\ref{AzumaY}) but do not employ the explicit form of 
${\hat D}_{s_{\rm x}}$. Therefore, the relationships 
such as 
$Y_{s_{\rm x},1_{\rm x}}=Y_{s_{\rm x},0_{\rm z}}+Y_{s_{\rm x},1_{\rm z}}-Y_{s_{\rm x},0_{\rm x}}$ that are considered
in the main text can be interpreted 
as the linear relationships between the $l^{th}$ {\it conditional} probabilities $Y_{s_{\rm x}, c|O_{l-1}}$. This is so because
the normalization factor $p^{(l)}$ does not affect this interrelation.
Thus, by taking the summation of such probabilities over $l$, Azuma's inequality~\cite{Azuma1, Azuma2} gives 
the actual occurrence number of such events and the phase error rate in the virtual protocol can be estimated.
This concludes the proof.

\section{Imperfect state preparation}\label{AP:four-state}
In this section we apply our phase error rate estimation technique to both a tilted four-state protocol, which 
is a
variant of the BB84 scheme, and the three-state protocol with modulation errors. 
For the former, we assume that the 
terminal points of the four Bloch vectors associated with the four states sent by Alice
form a triangular pyramid.

\subsection{A tilted four-state protocol}
 
Here we show that is possible to obtain the precise detection rate of any state. 
Suppose that Alice sends Bob four states given by
\begin{eqnarray}
{\hat \rho}_{j\alpha}=\frac{1}{2}\left({\hat \openone}+\sum_{t=x,y,z}p_{t}^{j\alpha}{\hat \sigma_{t}}\right),
\label{repre}
\end{eqnarray}
where $j\in\{0,1\}$ and $\alpha\in\{{\rm X,Z}\}$. 
Moreover, let us assume that the vectors ${\vec V_{j\alpha}}$, 
with ${\vec V_{j\alpha}}=(1,p_{x}^{j\alpha}, p_{y}^{j\alpha}, p_{z}^{j\alpha})$, are mutually linearly independent. From the viewpoint
of the Bloch sphere, this means that the terminal points of the four Bloch vectors associated with the four states form a triangular pyramid.
Suppose also that Alice and Bob distill key from the Z basis and use the events where Bob
employs the X basis for parameter estimation. 
In this scenario, let $\ket{\phi_{{j{\rm z}}}}_{{\rm A}_{e},{\rm B}}$ denote a purification of ${\hat \rho}_{j{\rm z}}$, with 
${\rm A}_{e}$ and $\rm B$ representing, respectively, Alice's shield system and the system that is sent to Bob. 
With this notation, Alice's state preparation process in the Z basis can be described by using any of the 
following two source states, 
\begin{eqnarray}
\ket{\Psi_{\rm Z}}_{{\rm A}, {\rm A}_{e}, {\rm B}}&=&\frac{1}{\sqrt{2}}\sum_{j=0,1}\ket{j_{\rm z}}_{\rm A}\ket{\phi_{j{\rm z}}}_{{\rm A}_{e}, {\rm B}},\label{unitary2} \\
\ket{\Psi_{\rm Z}}_{{\rm A}, {\rm A}_{e}, {\rm B}}&=&\frac{1}{\sqrt{2}}\sum_{j=0,1}\ket{j_{\rm z}}_{\rm A}\ket{\phi_{(j\oplus1){\rm z}}}_{{\rm A}_{e}, {\rm B}},\label{unitary3}
\end{eqnarray}
where $\rm A$ is a virtual qubit system of Alice and the symbol~$\oplus$ denotes the modulo-$2$ addition. 
If Alice measures system $\rm A$ in the Z basis she prepares the desire state at site $\rm B$, while 
she keeps the shield system ${\rm A}_{e}$. Here, the difference between Eqs.~(\ref{unitary2}) and (\ref{unitary3}) is
just a bit-flip. Since a bit-flip is a symmetry in
the problem, Alice is allowed to choose any of the two
equations above [Eqs.~(\ref{unitary2}) and (\ref{unitary3})] in constructing the purifications
with the goal of optimising the key generation rate.

To calculate the phase error rate $e_{\rm x}$ we consider the virtual protocol where 
Alice and Bob measure $\ket{\Psi_{\rm Z}}_{{\rm A}, {\rm A}_{e}, {\rm B}}$ in the X basis. 
In this virtual protocol, Alice emits
\begin{displaymath}
{\hat \sigma}_{{\rm B}; j_{\rm x}, {\rm Vir}}
={\rm Tr}_{{\rm A}, {\rm A}_{e}}\left[{\hat P}\left(\ket{j_{\rm x}}_{\rm A}\right)\otimes
{\hat \openone}_{{\rm A}_{e}, {\rm B}}{\hat P}\left(\ket{\Psi_{\rm Z}}_{{\rm A}, {\rm A}_{e}, {\rm B}}\right)\right].
\end{displaymath}
We denote these signals ${\hat \sigma}_{{\rm B}; j_{\rm x}, {\rm Vir}}$ as virtual states,
and we
define 
the normalised state ${\hat {\tilde \sigma}}_{{\rm B}; j_{\rm x}, {\rm Vir}}={\hat \sigma}_{{\rm B}; j_{\rm x}, {\rm Vir}}/{\rm Tr}\left({\hat \sigma}_{{\rm B}; j_{\rm x}, {\rm Vir}}\right)$.
Then, the joint probability that Alice sends $\ket{j_{\rm x}}$ and Bob detects the bit value $s_{\rm x}$ is given by
$Y_{s_{\rm x}, j_{\rm x}}=P(j_{\rm x}){\rm Tr}({\hat D}_{s_{\rm x}}{\hat {\tilde \sigma}}_{{\rm B}; j_{\rm x}, {\rm Vir}})$. 
Note that the value of $P(j_{\alpha})$, which is defined in the same way as $P(c)$ in the previous
section, is known from the protocol as well as Eqs.~(\ref{unitary2}) and (\ref{unitary3}). 
The phase error rate $e_{\rm x}$ is given by
\begin{equation} 
e_{\rm x}=\frac{Y_{0_{\rm x}, 1_{\rm x}}+Y_{1_{\rm x}, 0_{\rm x}}}{Y_{0_{\rm x}, 0_{\rm x}}
+Y_{1_{\rm x}, 0_{\rm x}}+Y_{0_{\rm x}, 1_{\rm x}}+Y_{1_{\rm x}, 1_{\rm x}}}.
\end{equation} 
Now, since ${\hat {\tilde \sigma}}_{{\rm B}; j_{\rm x}, {\rm Vir}}$ can also be written as
\begin{eqnarray}
{\hat {\tilde \sigma}}_{{\rm B}; j_{\rm x}, {\rm Vir}}=\frac{1}{2}\left({\hat \openone}+\sum_{t=x,y,z}p_{t}^{j_{\rm x},{\rm Vir}}{\hat \sigma_{t}}\right),
\label{repre2}
\end{eqnarray}
to obtain $Y_{s_{\rm x}, j_{\rm x}}$ (and thus $e_{\rm x}$) it is enough to calculate 
$q_{s_{\rm x}|t}={\rm Tr}({\hat D}_{s_{\rm x}}{\hat \sigma_{t}})/2$ with $t\in\{{\rm Id},x,y,z\}$.
For this, note that in the actual experiment 
we have the following constraints,
\begin{eqnarray}
Y_{s_{\rm x}, j_{\alpha}}&=&P(j_{\alpha}){\rm Tr}\left({\hat D}_{{s}_{\rm x}}{\hat \rho}_{j\alpha}\right)\nonumber\\
&=&P(j_{\alpha})\left(q_{s_{\rm x}|{\rm Id}}+p_{x}^{j\alpha}q_{s_{\rm x}|x}+p_{y}^{j\alpha}q_{s_{\rm x}|y}+p_{z}^{j\alpha}q_{s_{\rm x}|z}\right)\,.\nonumber\\
\label{appendLinear}
\end{eqnarray}
Then, as long as the vectors ${\vec V_{j\alpha}}$ are mutually linearly independent, we can solve 
the set of linear equations given by Eq.~(\ref{appendLinear})
and obtain $q_{s_{\rm x}|{\rm Id}}$, $q_{s_{\rm x}|x}$, $q_{s_{\rm x}|y}$, and $q_{s_{\rm x}|z}$. That is, 
we can determine the exact transmission rate of {\it any} state, including the signal ${\hat \sigma}_{{\rm B}; j_{\rm x}, {\rm Vir}}$, 
which gives the phase error rate. 

Moreover, if Bob's 
POVM elements act on a qubit space and he performs a measurement in the 
Y-basis, Alice and Bob can also estimate the Y-basis error rate. To see this, 
note that one only needs to change in the discussion above 
the terms $s_{\rm x}$ with $s_{\rm y}$ and define the Y-basis virtual state as
\begin{displaymath}
{\hat \sigma}_{{\rm B}; j_{\rm y}, {\rm Vir}}
={\rm Tr}_{{\rm A}, {\rm A}_{e}}\left[{\hat P}\left(\ket{j_{\rm y}}_{\rm A}\right)\otimes
{\hat \openone}_{{\rm A}_{e}, {\rm B}}{\hat P}\left(\ket{\Psi_{\rm Z}}_{{\rm A}, {\rm A}_{e}, {\rm B}}\right)\right].
\end{displaymath}

\subsection{Three-state protocol}

The technique described above can also be applied to the three-state protocol with modulation errors. 

In particular, suppose that the corresponding vectors ${\vec V_{j\alpha}}$ are mutually linearly independent and, 
moreover, 
the actual three states lie on the X-Z plane of the Bloch sphere. 
In this situation, we can consider the purifications 
given by Eq. (\ref{unitary2}) or Eq. (\ref{unitary3}) such that the coefficient $p_{y}^{j_{\rm x},{\rm Vir}}=0$ in Eq. (\ref{repre2}).
That is, all the 
coefficients of the purifications can be chosen to be real numbers in the X and Z bases. 
Now, since all the states, 
including the actual states and the virtual states, lie on the X-Z plane, we can obtain the transmission rate of the 
virtual states (and, consequently, the phase error rate $e_{\rm x}$) by just using the same arguments provided in the main text. 

This also implies that sending {\it any} three states, which do not necessarily lie on the X-Z plane, is enough to perform secure key distribution as long as 
the corresponding three vectors ${\vec V_{j\alpha}}$ are mutually linearly independent. That is, the terminal points of the Bloch vectors associated 
with the three states form a triangle.  
This is so because all the states on the X-Z plane can be uniformly ``lifted up" by a filtering operation 
$q\ket{0_{\rm y}}\bra{0_{\rm y}}+(1-q)\ket{1_{\rm y}}\bra{1_{\rm y}}$ with $0\le q<1$. That is, the virtual states 
can always be chosen on the same plane spanned by the three actual states. Thus, by using the transmission rate of
the actual states, one can obtain the transmission rate of the virtual states and, therefore, also the phase error rate.

\section{Simulation for the three-state protocol}
In this section we present the calculations used to obtain 
Fig.~1 (a) in the main text. We begin with the 
single-photon components of the signals sent by Alice. In particular, we have that the 
single photon part of 
$\ket{e^{i\xi}\sqrt{\alpha}}_{r}\ket{e^{i(\xi+\theta_{\rm A}+\delta\theta_{\rm A}/\pi)}\sqrt{\alpha}}_{s}$ is given by $(\ket{1}_{r}\ket{0}_{s}+e^{i (\theta_{\rm A}+\delta\theta_{\rm A}/\pi)}\ket{0}_{r}\ket{1}_{s})/\sqrt{2}$, 
where $0$ and $1$ represent, 
respectively, the photon number. The set 
$\{\ket{1}_{r}\ket{0}_{s}, \ket{0}_{r}\ket{1}_{s}\}$ forms a qubit basis. Therefore, 
we can choose the Z basis such that Alice's Z-basis states are expressed as
\begin{eqnarray}
\ket{\phi_{0{\rm z}}}&=&\ket{0_{\rm z}},\nonumber\\
\ket{\phi_{1{\rm z}}}&=&\sin\frac{\delta}{2}\ket{0_{\rm z}}+\cos\frac{\delta}{2}\ket{1_{\rm z}},
\end{eqnarray}
where $\ket{0_{\rm y}}:=\ket{1}_{r}\ket{0}_{s}$ and $\ket{1_{\rm y}}:=\ket{0}_{r}\ket{1}_{s}$ with 
$\ket{0_{\rm z}}:=(\ket{0_{\rm y}}+\ket{1_{\rm y}})/\sqrt{2}$ 
and $\ket{1_{\rm z}}:=(-i\ket{0_{\rm y}}+i\ket{1_{\rm y}})/\sqrt{2}$.

In the virtual protocol, Alice generates the state 
$(\ket{0_{\rm z}}_{\rm A}\ket{\phi_{0{\rm z}}}_{\rm B}+\ket{1_{\rm z}}_{\rm A}\ket{\phi_{1{\rm z}}}_{\rm B})/\sqrt{2}$ 
and sends system B to Bob. This joint state can be equivalently
expressed as 
\begin{equation}
\sqrt{1+\sin\frac{\delta}{2}}\ket{0_{\rm x}}_{\rm A}\ket{\phi_{0{\rm x}}'}_{\rm B}+\sqrt{1-\sin\frac{\delta}{2}}\ket{1_{\rm x}}_{\rm A}\ket{\phi_{1{\rm x}}'}_{\rm B},
\end{equation}
where the signals $\ket{\phi_{0{\rm x}}'}_{\rm B}$ and $\ket{\phi_{1{\rm x}}'}_{\rm B}$ have the form
\begin{eqnarray}
\ket{\phi_{0{\rm x}}'}_{\rm B}&:=&C_{0,0}(\delta)\ket{0_{\rm x}}_{\rm B}+C_{1,0}(\delta)\ket{1_{\rm x}}_{\rm B}\\
\ket{\phi_{1{\rm x}}'}_{\rm B}&:=&C_{0,1}(\delta)\ket{0_{\rm x}}_{\rm B}+C_{1,1}(\delta)\ket{1_{\rm x}}_{\rm B},
\end{eqnarray}
and the coefficients $C_{i,j}(\delta)$ are given by
\begin{eqnarray}
C_{0,0}(\delta)&=&\frac{1+\sin\frac{\delta}{2}+\cos\frac{\delta}{2}}{2\sqrt{1+\sin\frac{\delta}{2}}},
C_{1,0}(\delta)=\frac{1+\sin\frac{\delta}{2}-\cos\frac{\delta}{2}}{2\sqrt{1+\sin\frac{\delta}{2}}},\nonumber\\
C_{0,1}(\delta)&=&\frac{1-\sin\frac{\delta}{2}-\cos\frac{\delta}{2}}{2\sqrt{1-\sin\frac{\delta}{2}}},
C_{1,1}(\delta)=\frac{1-\sin\frac{\delta}{2}+\cos\frac{\delta}{2}}{2\sqrt{1-\sin\frac{\delta}{2}}}. \nonumber
\end{eqnarray}
As already explained in the main text, 
our phase error rate estimation technique 
provides the exact value for the transmission rates 
$Y_{s_{\rm x},j_{\rm x}}$.
Therefore, we can obtain the precise transmission rate of the signals $\ket{\phi_{j{\rm x}}'}_{\rm B}$. 

For our simulations, we consider a
channel model where the conditional probabilities $Y_{s_{\rm x}|j_{\rm x}}$ 
({\it i.e.}, the conditional probability that Bob obtains 
$s_{\rm x}$ given that Alice sent him the state $\ket{j_{\rm x}}$) are given by
\begin{eqnarray}\label{channel}
Y_{s_{\rm x}|j_{\rm x}}&=&(1-L)C_{s,j}(3\delta/2)^{2}(1-e_{\rm d}/2)+e_{\rm d}(1-e_{\rm d}/2)\nonumber\\
&+&(1-L)C_{s\oplus1, j}(3\delta/2)^{2}e_{\rm d},
\end{eqnarray}
where $e_{d}$ is the dark count rate of Bob's detectors, and $L$ denotes the total loss rate. 
Here, we have considered the transformation $\delta\rightarrow3\delta/2$ because Bob applies a phase modulation
to the incoming signals. In Eq.~(\ref{channel}), 
the first (second) term models a single detection click at Bob's side produced by a photon (dark count), while the last term 
represents simultaneous clicks. Note that in this last case (simultaneous clicks), 
Bob assigns a random bit value to the measurement result. 

For convenience, here we will write the phase error rate 
$e_{\rm x}$
in terms of the conditional
probabilities $Y_{s_{\rm x}|j_{\rm x}}$. Note that we can do so because the choice of the state and
the measurement is random and uniform. We have that
the phase error rate $e_{\rm x}^{(1)}$ of the single-photon components, and
the single-photon gain $Q_{\rm {\rm z}}^{(1)}$ are given by
\begin{eqnarray}
e_{\rm x}^{(1)}&=&\frac{Y_{1_{\rm x}|0_{\rm x}}+Y_{0_{\rm x}|1_{\rm x}}}{Y_{1_{\rm x}|0_{\rm x}}+Y_{0_{\rm x}|1_{\rm x}}+Y_{1_{\rm x}|1_{\rm x}}+Y_{0_{\rm x}|0_{\rm x}}}\,,\\
Q_{\rm z}^{(1)}&=&\frac{1}{2}e^{-2\alpha}\alpha\sum_{j,s}Y_{s_{\rm x}|j_{\rm x}}P(j_{\rm x}),
\end{eqnarray}
where $P(j_{\rm x})=[1+(-1)^{j}\sin{(\delta/2)}]/2$.

 Similarly, one can obtain the overall 
gain $Q_{\rm z}$ and the bit error rate $e_{\rm z}$ in the Z basis. These parameters have the form
\begin{eqnarray}
e_{\rm z}&=&w_{\rm z}/Q_{\rm z}\,,\\
Q_{\rm z}&=&\frac{1}{2}\left[P_{0|0}(1-P_{1|0})+(1-P_{0|0})P_{1|0}+P_{0|0}P_{1|0}\right]\nonumber\\
&+&\frac{1}{2}\left[P_{0|1}(1-P_{1|1})+(1-P_{0|1})P_{1|1}+P_{0|1}P_{1|1}\right], \nonumber\\ \\
w_{\rm z}&=&\frac{1}{2}\left[P_{0|0}(1-P_{1|0})+(1-P_{0|0})P_{1|0}+P_{0|0}P_{1|0}\right],\nonumber\\
\end{eqnarray}
where $P_{s|j}$ is the conditional probability that Bob obtains a bit value 
$s$ given that Alice sent him a bit value $j$. These probabilities can be written as 
\begin{eqnarray}
P_{0|0}&=&e_{\rm d}+(1-e_{\rm d})\left[1-e^{-\alpha(1-L)}\right],\\
P_{1|0}&=&e_{\rm d},\\
P_{0|1}&=&e_{\rm d}+(1-e_{\rm d})\left[1-e^{-\alpha(1-L)\sin^2{(\delta/2)}}\right],\\
P_{1|1}&=&e_{\rm d}+(1-e_{\rm d})\left[1-e^{-\alpha(1-L)\cos^2{(\delta/2)}}\right].
\end{eqnarray}
Finally, the asymptotic key generation rate is given by
\begin{eqnarray}
R=\frac{1}{2}\left\{Q_{\rm z}^{(1)}\left[1-h(e_{\rm x})\right]-Q_{\rm z}h(e_{\rm z})\right\}\,,
\end{eqnarray}
where $h(x)$ is the binary entropy function. 
For each value of the distance, we optimise the parameter
$\alpha$ to maximise the key rate. The result is 
shown in Fig.~1 (a) in the main text.

\bibliographystyle{apsrev}

\end{document}